%% file: 1-vis-template.tex
\documentclass[preprint,journal]{vgtc}            % preprint (journal style)

\onlineid{1129}

\vgtccategory{Research}

\vgtcpapertype{Empirical}

\title{Designing Within the Lines: Practitioners' Perspectives and Visualisation Tool Evaluation in the Arabic Context}

\author{%
  \authororcid{Muna Alebri$^{1}$}{0000-0002-6471-6206},
  \authororcid{Noëlle Rakotondravony$^{1}$}{0000-0002-7594-6349}, 
 \authororcid{Yassine Bechqito}{0009-0009-7946-0459}, 
  \authororcid{Georgia Panagiotidou}{0000-0003-4408-6371}, \\
  \authororcid{Lane Harrison}{0000-0003-3029-2799}, 
  \authororcid{Hassan Aldhanhabi}{}, and 
  \authororcid{Salem Alkaabi}{} 
}

\authorfooter{
\item
  	MA$^{1}$, YB, HA, and SA are with United Arab Emirates University, UAE. 
    E-mail: \{munaalebri, ybechqito, 202020708, 202000336\}@uaeu.ac.ae
\item
  	NR$^{1}$ and LH are with Worcester Polytechnic Institute.
  	E-mail: \{ntrakotondravony, ltharrison\}@wpi.edu
\item
  	GP is with King's College London.
  	E-mail: georgia.panagiotidou@kcl.ac.uk
\item[$^{1}$]Both authors contributed equally to this work
}

\abstract{
Design guidelines and best practices serve as references that support designers throughout the visualisation design process.
While considerable effort has identified the elements that contribute to effective data visualisations, little attention has been paid to how language (scripts and reading direction), tool support, and cultural context also shape design decisions. 
As a result, assumptions of homogeneity persist, with visualisation practices predominantly benefiting users of English and left-to-right (LTR) scripts while overlooking the needs of over two billion Arabic script users.
We investigate how Arabic-speaking visualisation practitioners design for right-to-left (RTL) scripts. We report on an analytical evaluation of seven popular GUI-based visualisation authoring tools using an Arabic dataset %with both Western and Eastern numerals, 
complemented by interviews with 11 Arabic-speaking practitioners across journalism, design, and data analysis.
Our findings reveal that visualisation practitioners constantly negotiate tensions between Arabic reading conventions, ``universal'' LTR visual norms, and limited tool support for Arabic text, Eastern numerals, and maps. 
They engage in substantial labour, such as manually mirroring charts, fixing alignment issues, and stitching together multi-tool workflows, while making strategic compromises in language choice, interactivity, and chart type. 
Our tool analysis further reveals fragmented, inconsistent support for RTL mirroring, poor numeral rendering, and map defaults that encode geopolitical assumptions. 
In light of these findings, we discuss how RTL visualisation work is carried out under many constraints that affect agency and creativity. 
We argue that visualisation tools and defaults operationalise linguistic and geopolitical power in RTL contexts, and offer research directions and design implications that more robustly support RTL practitioners. 
Supplementary material: \href{https://osf.io/8esbf/overview?view_only=f79ab947ed9c4f0489bcf2a02c3c8514}{https://osf.io/8esbf}.
}

\keywords{Visualizations, Arabic Visualizations, Right to Left, Design Practice, Tool Comparison}

\teaser{
  \centering
  \includegraphics[angle=0, width=\linewidth, alt={An overview of the challenges faced by practitioners using visualisation authoring tools with Arabic scripts (left panel) and how they address these challenges (right panel). 
  These are based on an analytical evaluation of seven visualisation tools and interviews with 11 Arab practitioners.
  Each panel include quotes from interviewed participants, illustrating how the challenges affect their design works, and how they address them.}]{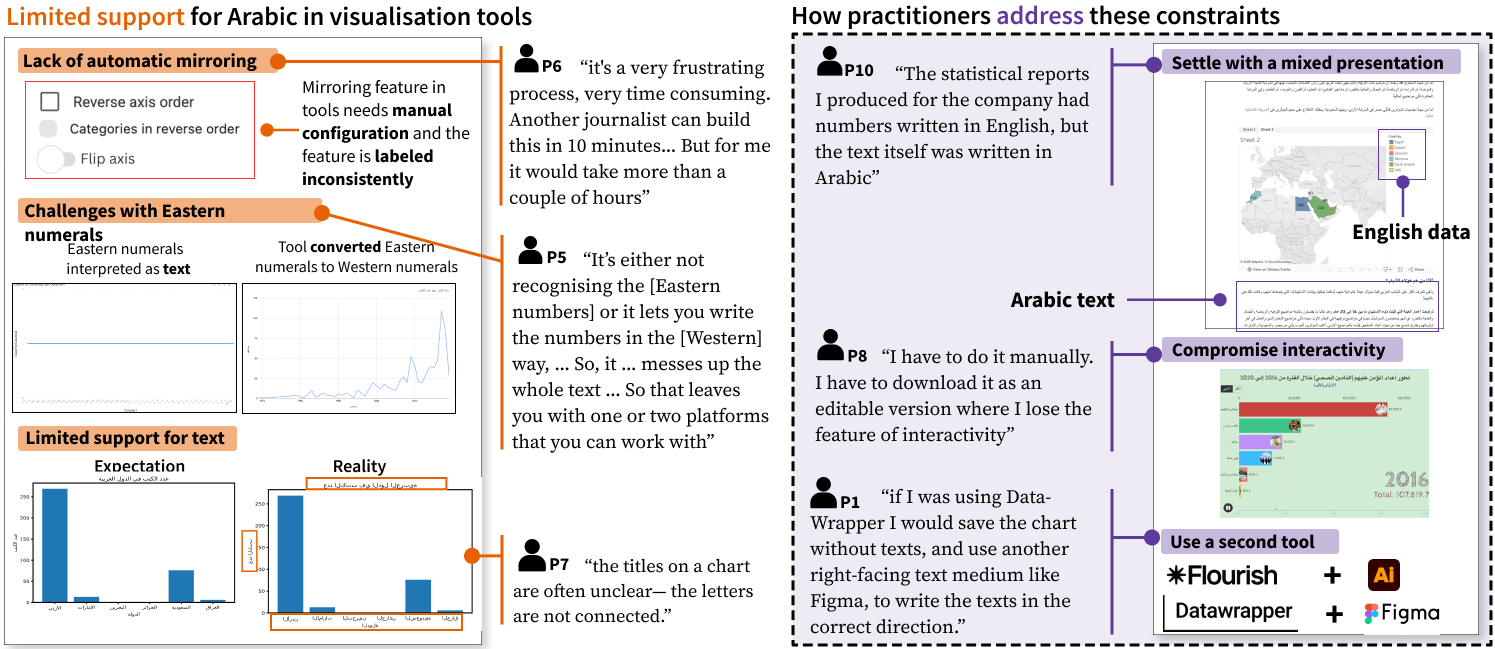}
  \caption{
  	A snapshot of the challenges faced by practitioners using visualisation authoring tools with Arabic scripts and how they address them. These are based on an analytical evaluation of seven visualisation tools and interviews with 11 Arab practitioners. 
  }
  \label{fig:teaser}
}

\graphicspath{{figs/}{figures/}{pictures/}{images/}{./}} % where to search for the images

\usepackage[bidi=default, english]{babel}
\usepackage[LAE, T1]{fontenc}

\usepackage{tabu}                      % only used for the table example
\usepackage{booktabs}                  % only used for the table example
\usepackage{lipsum}                    % used to generate placeholder text
\usepackage{mwe}                       % used to generate placeholder figures

\usepackage[table]{xcolor}
\usepackage[x11names]{xcolor}
\usepackage{array}
\usepackage{graphicx}

\usepackage{mathptmx}                  % use matching math font
\usepackage{tikz}
\definecolor{lightpeach}{RGB}{255,218,185}
\usepackage{comment}

\usepackage{enumitem} % to be able to use \begin{itemize}[noitemsep]
\usepackage{xspace}
\usepackage{soul} % for highlight \hl command

\usepackage{tabularx}
\usepackage{amssymb}

\usepackage{censor}
\usepackage[most]{tcolorbox} % for highlights' round corners

\usetikzlibrary{calc}
\newcommand*\circled[1]{\tikz[baseline=(char.base)]{
    \node[shape=circle, draw, inner sep=1pt, 
        minimum height={\f@size*1.6},] (char) {\vphantom{WAH1g}#1};}}
\makeatother

\newcommand{\etal}{\emph{et al.}\@\xspace}

\newcommand{\eg}{\emph{e.g.}\xspace}

\newcommand{\etals}{\mbox{\emph{et~al.}'s }}

\definecolor{quotecolor}{HTML}{273372}

\newcommand{\checkblack}{{\color{black}$\checkmark$}}

\definecolor{myblue}{RGB}{31,111,160}

\definecolor{lightbluecustom}{RGB}{204,232,255}

\newenvironment{bluequote}
  {\begin{quote}\color{myblue}}%
  {\end{quote}}

\definecolor{limitation}{HTML}{E66100} % colorblind safe colors from https://davidmathlogic.com/colorblind/#%23E66100-%235D3A9B
\newcommand{\hlred}[1]{
\unskip
  \tcbox[on line, colback=limitation!30, colframe=limitation!30,
    arc=0.5pt, boxsep=0pt, left=0pt, right=0pt, top=0pt, bottom=0pt]{#1}}

\definecolor{myLightBlue}{RGB}{212,226,252}
\definecolor{address}{HTML}{5D3A9B}
\newcommand{\hlblue}[1]{
\unskip
  \tcbox[on line, colback=address!30, colframe=address!30,
    arc=0.5pt, boxsep=0pt, left=0pt, right=0pt, top=0pt, bottom=0pt]{#1}}

\definecolor{LightPurple}{RGB}{243,209,246}

\definecolor{LightGreen}{RGB}{199, 246, 199}

\definecolor{LightYellow}{RGB}{222, 184, 135}

\definecolor{LightOrange}{RGB}{146, 189, 255}
\newcommand{\hlLightOrange}[1]{\xspace
  \tcbox[on line, colback=LightOrange, colframe=LightOrange,
    arc=0.5pt, boxsep=-0pt, left=0pt, right=0pt, top=0pt, bottom=0pt]{#1}}

\definecolor{ThistleOne}{RGB}{255,215,255}
\newcommand{\hlthistleone}[1]{
  \tcbox[on line, colback=ThistleOne, colframe=ThistleOne,
    arc=0.5pt, boxsep=-0pt, left=0pt, right=0pt, top=0pt, bottom=0pt]{#1}}

\definecolor{DarkSeaGreenOne}{RGB}{193,255,193}
\newcommand{\hldarkseagreen}[1]{
  \unskip
  \tcbox[on line, colback=DarkSeaGreenOne, colframe=DarkSeaGreenOne,
    arc=0.5pt, boxsep=-0pt, left=0pt, right=0pt, top=0pt, bottom=0pt]{#1}}

\usepackage{quoting}
\newcommand*{\blockquote}[1]{%
  \begin{quoting}[ indentfirst = false,
    noorphans = true, leftmargin=0.5em, rightmargin=0em, vskip=0em]
  \par\addvspace{0.1em}
    #1
    \par\addvspace{0.2em}
  \end{quoting}}

\newcommand{\ourquote}[1]{ \textcolor{quotecolor}{\textit{#1}}}

\newcommand{\aquote}[1]{\textcolor{quotecolor}{\textit{#1}}}
\newcommand{\alongquote}[2]{\blockquote{\ourquote{#1} (#2)}}

\usepackage[normalem]{ulem}
\usepackage{xcolor}

\begin{document}

%%%%%%%%%%%%%%%%%%%%%%%%%%%%%%%%%%%%%%%%%%%%%%%%%%%%%%%%%%%%%%%%
%%%%%%%%%%%%%%%%%%%%%% START OF THE PAPER %%%%%%%%%%%%%%%%%%%%%%
%%%%%%%%%%%%%%%%%%%%%%%%%%%%%%%%%%%%%%%%%%%%%%%%%%%%%%%%%%%%%%%%

%% The ``\maketitle'' command must be the first command after the
%% ``\begin{document}'' command. It prepares and prints the title block.
%% the only exception to this rule is the \firstsection command
% \firstsection{Introduction}

\maketitle

\input{rtl-main}
% \input{vis-2026-template/2-vis-main}

%% if specified like this the section will be omitted in review mode
\acknowledgments{%
	This work was supported in part by the National Science Foundation \#2213757.%
}

\bibliographystyle{abbrv-doi-hyperref}

\bibliography{1-vis-template}

\appendix % You can use the `hideappendix` class option to skip everything after \appendix

\end{document}

%% file: rtl-main.tex
\newcolumntype{Y}{>{\centering\arraybackslash}X}

%################################ SECTION ####################
\section{Introduction}

The widespread adoption of technology has expanded the number and diversity of users who encounter, interpret, and rely on visualisations every day. 
Yet, visualisation research predominantly focused on users from North America \cite{Jena2021TheVisualization}. 
Therefore, insights from empirical research may not apply to non-WEIRD (Western, educated, industrialised, rich, and democratic) populations \cite{Henrich2010MostWEIRD}. Considering only WEIRD populations leads to treating English as the default, reinforcing homogeneity and obscuring diversity. While English accounts for more than 60\% of internet content \cite{Chua2022HowWorld}, other scripts, such as Arabic, written from right-to-left (RTL), are also widely used, with over two billion users \cite{w32023languages}. 

\begin{comment}
Designing effective visualisations has been a central objective for both practitioners and researchers. 
Scholars broadly recommend effective design strategies based on empirical evidence, such as making a visualisation memorable \cite{Borkin2013WhatMemorable} and persuasive \cite{Garreton2023AttitudinalStories}. 
Yet, these recommendations often give limited attention to language and its directionality, as well as to culture. 
In fact, most contributions on visualisation design practice are based on samples from North America and Europe, a limitation that many papers acknowledge (e.g., \cite{Parsons2022UnderstandingPractice, Jena2021TheVisualization}). 
Overlooking language and culture leads to treating English as the default, reinforcing homogeneity and obscuring diversity.
\end{comment}

The importance of considering culture, language, and its directionality lies in fundamental differences in perception and preference. 
Arunkumar \etal \cite{arunkumar2025lost} show that bilingual users prefer English annotations in a chart for surgical precision and Native-language annotations for holistic understanding. 
In colour, Gibson \etal \cite{Gibson2017ColorUse} showed that colour names vary across cultures due to communication needs and industrialisation. 
Similarly, Afsari \etal \cite{Afsari03072018} found that RTL script users varied in their viewing behaviour, while those who use left-to-right (LTR) scripts showed stability and bias to viewing from the left. 

%While English accounts for more than 60\% of internet content \cite{Chua2022HowWorld}, other scripts, such as Arabic, which is written from RTL, are also widely used, with over two billion users \cite{w32023languages}. 
The issue extends beyond linguistic variation to encompass fundamental visuo-spatial orientations shaped by script direction \cite{Fagard2003TheChildren, fuhrman2010crosscultural}. 
Prior work \cite{alebri2024design} mapping the visualisation design space in Arabic data-driven articles has demonstrated ad-hoc adaptations, but these practices remain inconsistent and rarely grounded in empirical evidence.
This inconsistency might also stem from how the visualisation tools support the designers' work. 
Visualisation authoring tools are powerful; prior research \cite{Parsons2020DataChartjunk} shows that visualisation designers tend to avoid using embellishments, even when aware of their benefits for their audiences, due to a lack of appropriate tools. The lack of characterisation of the extent to which widely used visualisation authoring tools support Arabic script, Eastern numerals, bidirectional text, and RTL mirroring remains unclear, particularly on how it might drive RTL design decisions in practice. In this work, we pose the following research questions:

\begin{itemize}[topsep=0.5pt,  ]
    \item RQ1: How do visualisation practitioners navigate design decisions when creating Arabic visualisations?
    \item RQ2: How do tool features or guidelines support them in addressing script-specific and cultural challenges during the design process?
\end{itemize}

To address these questions, we conducted two complementary investigations.
First, using an Arabic dataset \cite{alharthi2021jamalon}, we evaluated seven popular Graphical User Interface (GUI)-based visualisation authoring tools: Microsoft Excel, Tableau, Power BI, Flourish, Datawrapper, Google Sheets, and RAWGraphs. Given the identified limitations of the tools, we interviewed 11 Arabic-speaking visualisation practitioners across journalism, design, and data analysis.
We examined how they reason about directionality, navigate tool limitations, and what are the processes behind their design decisions.

Our results reveal shortcomings and inconsistencies in the visualisation tools' support for Arabic scripts and RTL design. 
While some tools provide partial mirroring options, these are unevenly implemented and labelled; Eastern numerals are frequently misrecognised as text, and map defaults sometimes alter location labels and boundaries.
These limitations stem partially from the lack of control granted to the practitioner and require additional time and effort on foundational design elements, %(\eg, numerical systems, alignment)
thereby compromising their creativity. 
The tools we analysed were predominantly designed for English and follow a LTR convention. In fact, even while writing this manuscript, we encountered challenges in including Arabic text within the template, suggesting a deviation from the norm.
We also observed that the interviewees themselves reinforced the stance of homogeneity by assuming their audiences' data visualisation literacy and habitual reading from their own experiences.

We reflect on our findings in light of prior work on visualisation conventions and the rhetorical use of visualisations, and argue that there is an infrastructural asymmetry, demonstrating how defaults in visualisation tools privilege LTR conventions, along with calls for action.
By highlighting the interplay between cultural contexts and technical infrastructures, our work not only benefits Arab practitioners but also contributes to broader inclusivity.

%################################ SECTION ####################
\section{Background}
We review the literature on visualisation design practice in relation to RTL scripts, the impact of language and culture on visualisation interpretation, and the state of visualisation authoring tools. 

% For RQ1 and consequently RQ2?
\subsection{RTL Design and Visualisations in Practice} 
% \hl{Designing effective visualisations has been a central objective for both practitioners and researchers.}
The effectiveness of data visualisations hinges on the design practices which encompass not only the technical execution, but also the series of deliberate design decisions.
These decisions help structure how data is presented, what to include or omit in the encoding, such as colour, chart type, annotations, etc. 
% Much work guiding design practices focus on practitioner guidelines \cite{Ajani2021DeclutterCommunication, Diehl2018VisGuides, ottley2026consensus}, evaluating visualisation authoring tools and resources \cite{huang2025comparison, liu2023visualization}. 
Prior work has mapped the landscape of visualisation design practices across domains, such as data-driven articles \cite{Hao2024DesignArticles}, anthropographics \cite{Morais2020ShowingAnthropographics}, thumbnail visualisations \cite{Kim2019ThumbnailsPractices}, and data comics \cite{Bach2018DesignComics}.
Other research has also contributed to our understanding of data visualisation practices.
%For example, Parsons \etal \cite{Parsons2020WhatUse} surveyed design principles that practitioners use or are familiar with. In another body of work, they also interviewed designers to understand their design process \cite{Parsons2022UnderstandingPractice} and the source of their inspiration \cite{Baigelenov2025HowInspiration}. 
For example, Parsons \etal \cite{Parsons2020WhatUse} conducted a survey where they identified design principles (\eg, data-ink ratio, cognitive load) that practitioners use or are familiar with. 
Similarly, interviews with practitioners helped understand their design process \cite{Parsons2022UnderstandingPractice} and the source of their inspiration \cite{Baigelenov2025HowInspiration}.
Today, modern visualisation design practices are shaped by widely adopted tools and practitioners' design guidelines (\eg, \cite{Ajani2021DeclutterCommunication, Diehl2018VisGuides, ottley2026consensus}). 
% Guidelines can sometimes expedite design decisions 
% % by drawing on researched and tested conventions or practical experience \cite{kelleher2011ten, kandogan2016grounded}, 
% and tools act as the infrastructure for operationalising design decisions through conventions, recommendations, and defaults.
% Despite these advances, language and culture are insufficiently addressed within the visualisation design practice. 
% Kandogan \etal identified \textit{users} as one of the five high-level concepts that characterize visualisation guidelines: "User (U): User entails all aspects related to human viewing and interacting with visual representation of data, including tasks (e.g. compare, communicate), (dis)abilities (e.g. read, recognize, recall), and social aspects (e.g. conventions) "\cite{kandogan2016grounded}.

However, factors such as the designer's language, their culture, and the cultural contexts in which they operate are rarely considered in understanding the visualisation design process, leaving open questions about how practices translate across contexts.
Research in adjacent fields like HCI and UI have extensively addressed cross-cultural adaptation through internationalisation (i18n) and localisation (l10n) \cite{marcus2009global, reinecke2011improving}, defining adjustments for RTL contexts such as text alignment, layout mirroring, and bidirectional text support \cite{marcus2009global, aykin2004usability}.
Industry standards (\eg, W3C \cite{w3structural}), and design systems (\eg, Google Material Design \cite{googlematerial}) also formalise these principles, although they primarily target user interfaces and navigation elements rather than the layout of data-driven graphs.
In data visualisation, applying these principles might not be trivial for designers, and requires coordination between tool support, design processes, and an understanding of RTL audiences' expectations for visual encodings of data. 
% But what happens when designers create visualisations for RTL audiences using tools and guidances built around LTR conventions?

In this work, we address this gap, asking what happens when creating visualisations for RTL audiences using tools and guidance grounded in LTR conventions? 
Focusing on Arabic and RTL contexts, we foreground the perspective of RTL data visualisation practitioners who come from understudied geographical areas, and design for understudied languages in the field of data visualisation.
\subsection{Reading direction and visualisation interpretation}

Reading direction is closely embedded in the perceptual process of how individuals scan and interpret data visualisation.
Research in cognitive and cross-cultural psychology shows that habitual reading/writing directions (\eg RTL, LTR) bias how people attend to, scan, and organise visual information \cite{Fagard2003TheChildren, abed1991cultural, fuhrman2010crosscultural}.
For example, Arabic and Hebrew readers tend to prefer rightward directions when judging motion, sequences, or temporal progress, compared to English readers \cite{maass2003directional, tversky1991cross}.
These effects or directional biases can be dynamic and learned. 
% and can be modulated by literacy, bilingualism, and task context, highlighting the interplay between culture and cognition \cite{guida2018spatialization}.
G\"obel \etal found that pre-literate children from LTR and RTL cultures exhibit different default counting directions, which can temporarily reverse after short shared-reading experiences \cite{gobel2018observation}.
This suggests that reading-direction conventions are culturally acquired and shaped by the context in which the materials are consumed. 
For data visualisations in RTL context, this challenges the universality of LTR assumptions in visualisation design and highlights the impact of layout direction on interpretation.

% Why direction matters
In data visualisation, reading direction habits and biases have been linked to how people interpret visual encodings. 
Directional biases influence where viewers attend first when scanning a visualisation \cite{segel2010narrative}, and common charts such as bar charts, timelines, and scatterplots tend to align with habitual reading flow, such that orientation and axis placement can impact both comprehension and perceived naturalness \cite{hullman2011visualization}. 
Cross-cultural research further shows that spatial metaphors such as ``progress”, ``increase”, or ``future” are interpreted differently across reading traditions \cite{fuhrman2010crosscultural}, suggesting a broader role of directionality in shaping how visual information is understood.

% While cross-cultural empirical visualization research is limited,
% Most empirical studies assumes LTR reading order, 
Our knowledge about human perception in visualisation is mostly based on LTR script users, 
% (\eg, \cite{Cleveland1984GraphicalMethods, Borkin2013WhatMemorable, fygenson2024arrangement})
with English as the dominant language of research and  pedagogy\cite{rakotondravony2023english}. 
Cross-cultural data visualisation research on Arabic or RTL contexts show that Arabic-speaking users have different interpretation and preferences with basic chart styles \cite{rakotondravony2023probablement, arunkumar2025lost, gorelik2025reading}, and designers rely on ad hoc strategies like mirroring, axis reversal, mixed directions to reconcile linguistic and graphical conventions \cite{alebri2024design}.
We investigate how creators of RTL data visualisation reason about these adaptations. 
% As prior work shows, the visual and textual elements of a design constitute complementary channels that readers integrate when forming takeaways \cite{stokes2022striking, hearst2023show}. 
Because visual elements and the language through text are interrelated, complementary channels that readers integrate together \cite{stokes2022striking, hearst2023show}, we approach these adaptations holistically. Understanding when creators align or separate linguistic and visual directions is key to advancing culturally inclusive visualisation practices.
\subsection{Evaluating Data Visualisation Authoring Tools} %Defaults, and Automation}
Data visualisation tools are central to the design process, motivating their evaluation across domains \cite{behrisch2018commercial, huang2025comparison, ravi2013survey, shakeel2022comprehensive}. 
As mainstream tools such as Tableau, Power BI, Microsoft Excel, and Google Sheets are widely used beyond scientific communication and among non-programmers \cite{Data2024DataSurvey}, examining their capabilities is key to understanding how they enable or constrain design practice.
Prior work has evaluated these tools by comparing their usability, functionality, and support for designers’ workflows. 
Common approaches involve applying a reference dataset across tools and analysing the resulting outputs or user processes.
For instance, Negi \etal \cite{negi2025charting} compared Tableau, Power BI, Looker Studio, and Qlik Sense for rendering large-scale and real-time data; Huang \etal \cite{huang2025comparison} examined how dashboard developers used Tableau, Spotfire, Power BI, and R Shiny for presenting scientific data in the health domain.
Other studies have also assessed usability \cite{negi2025charting}, support for interactivity and other visualisation techniques \cite{schmidt2020usage}, scalability \cite{pandey2022comparative}, and workflow integration \cite{lousa2019evaluation}.

These evaluations provide valuable insights into tool capabilities and inform practitioners’ choices based on organisational needs, data scale, and desired design outcomes\cite{huang2025comparison, lavanya2023comprehensive, behrisch2018commercial}. 
However, they largely foreground datasets and defaults aligned with LTR scripts, and thus rarely examine how popular tools support RTL visualisation design. 
Because tool defaults shape what is easy or difficult to produce, the limited visibility of Arabic-script and RTL support in features and settings remains a critical gap. 
% Although inconsistencies in RTL visualisation practices have been noted \cite{alebri2024design}, 
In this work, we address this gap and explore the relationship between the technical and functional affordances of tools and practitioners’ experiences in RTL contexts.
% Additionally, in UI design and engineering,internationalization and localization are often treated as merely technical concerns (\eg unicode support, locale detection, translation, etc...) \cite{esselink2000practical}, 

%################################ SECTION ####################
\section{Analytical Evaluation of Visualisation Tools}
\label{sec:tool-analysis}

\newcommand{\iconmap}{\includegraphics[height=1em]{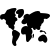}\xspace}
\newcommand{\iconbar}{\includegraphics[height=1em]{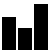}\xspace}
\newcommand{\iconline}{\includegraphics[height=1em]{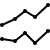}\xspace}

\begin{table*}[h]
\setlength{\belowrulesep}{3pt}

\centering
\caption{Codebook guide for evaluating the support for Arabic scripts in the analysed visualisation authoring tools. }
\label{tab:tool-codebook}
\begin{tabular}{ll>{\fontsize{8pt}{10pt}\selectfont}p{12.3cm}}
\toprule
 & \multicolumn{1}{l|}{Name} & \normalsize \textbf{Description}  \\ 
\midrule
\rowcolor{LightPurple}
\multicolumn{3}{l}{\textbf{Tool characteristics}} \\

 & \multicolumn{1}{l|}{Tool Arabic support}   & Does the tool exist in a version in Arabic? (y, n,y in premium)    \\
 & \multicolumn{1}{l|}{UI support for RTL}    & How does the UI switch to a RTL layout if the language is set to Arabic (\eg All mirrored, labels translated)\\%, No support) \\
 & \multicolumn{1}{l|}{Platform access}   & How can the tool be accessed ( Desktop, web-based, both)    \\
 & \multicolumn{1}{l|}{License}   & What are the available plans for the tool? (\eg  Commercial, Freemium, Open source, Free)  \\
 & \multicolumn{1}{l|}{Primary user group}    & Who are the primary users for the tool? (\eg  Data Analysts, Developers, General Public, Researchers, Others)   \\
 & \multicolumn{1}{l|}{Chart types supported} & The chart types that can be generated (\eg  BarChart, LineChart)   \\
 & \multicolumn{1}{l|}{RTL mirroring support} & Can chart be mirrored? (y-one step, y-multiple steps, n)    \\
 & \multicolumn{1}{l|}{Interactivity} & Does the tool allow creating interactive visualisations? (y/n)  \\
 & \multicolumn{1}{l|}{Calendar format}   & Does the tool support calendars other than the Gregorian calendar (e.g., Hijri, Hebrew)? (y/n)  \\
 & \multicolumn{1}{l|}{Font support}  & Fonts available for Arabic script (Included in defaults, Configurable, No support) \\ 
\midrule
\rowcolor{LightOrange} \multicolumn{3}{l}{\textbf{Community support}} \\ 
 & \multicolumn{1}{l|}{Arabic documentation availability} & Does the tool provide documentation in Arabic? (y/n) \\
 & \multicolumn{1}{l|}{Community support for RTL issues}  & Is there a community support for RTL issues? (Active, Somewhat active, Inactive, Does not exist) \\
 %& \multicolumn{1}{l|}{Custom plugin support} & Does the tool allow plug-ins/libraries to support Arabic scripts? (y/n) \\ 

\midrule
\rowcolor{DarkSeaGreenOne}
\multicolumn{3}{l}{\textbf{Default charts (barchart \iconbar, linechart \iconline, map \iconmap using Eastern numerals (\foreignlanguage{arabic}{١، ٢، ٣}) and Western numerals (1, 2, 3))}}  \\ 
 & \multicolumn{1}{l|}{Arabic text rendering} & Correctly render axis Arabic text? (y/n) \\
 & \multicolumn{1}{l|}{Bi-directional text handling}  & Can it handle AR text with EN text or numerals?  tested with writing a title (y/n)   \\
 & \multicolumn{1}{l|}{Eastern number rendering}  & Does the tool correctly render axis Eastern numbers? (y/n)   \\
 & \multicolumn{1}{l|}{Legend alignment}  & Where's the legend placed in maps? (\eg  Left-aligned, Right-aligned, Configurable, couldn't display (C/D)) \\
 & \multicolumn{1}{l|}{X-axis direction}  & What is the X-axis direction for the numerical axis? (LTR, RTL, Configurable, C/D, N/A) \\
 & \multicolumn{1}{l|}{Y-axis position}   & What is the Y-axis position in bar and linecharts?(Left, Right, Configurable, C/D, N/A) \\
 & \multicolumn{1}{l|}{Tooltip and annotation position}   & Where is the position of the tooltip? (\eg  Left-to-pointer, Right-to-pointer, Dynamic, Configurable, C/D, N/A)  \\ 
 \bottomrule
\end{tabular}
\end{table*}

To understand how existing visualisation authoring tools support RTL visualisations, we performed an analytical evaluation. This step is necessary not only to understand the tool's limitations, but also to comprehend the design decision process practitioners follow when designing Arabic visualisations. In the following sections, we describe our method for testing visualisation tools' support for Arabic scripts.

\subsection{Visualisation authoring tools}
We evaluated seven visualisation authoring tools on their support for Arabic scripts. 
We utilised the dataset published by Alebri \etal \cite{alebri2024design} where they identified that Flourish, Highcharts, and D3 have been used in generating some of the visualisations found in Arabic data-driven articles. 
We also based our selection on the most used tools identified through the state of the industry survey (which included designers, journalists, and data analysts) conducted by the Data Visualisation Society in 2024 \cite{Data2024DataSurvey}: Microsoft Excel, Tableau, Google Sheets, Python, Figma, R, Power BI, Adobe Illustrator, D3, Datawrapper, Flourish, and RAWGraphs. 
We focused on GUI-based tools and excluded code-based tools (e.g., Python libraries, R), tools that require plugins to generate charts (e.g., Figma), and tools primarily used for editing (e.g., Illustrator). %GUI-based tools also pre-encode default values and visualisation designs, and we aim to explore these presets. %as these typically offer less flexibility and pre-encode default values and visualisation designs. 
%Accordingly, we excluded frameworks and libraries such as Python libraries (\eg Matplotlib), D3, and Highcharts, as these enable the creation of highly customizable charts. 
Therefore, we evaluated: \textit{Microsoft Excel, Tableau, Power BI, Flourish, Datawrapper, Google Sheets, and RAWGraphs}. 

\subsection{Dataset} 

We used the Jamloon Arabic Book dataset \cite{alharthi2021jamalon} %, which contains more than 8000 titles, 
 to evaluate the visualisation tools' support for RTL scripts. We selected this dataset as it contained time-series, numerical, and categorical data, and the textual data was in Arabic. For geographical data, we extracted the country publication from the publisher column.% and assigned longitude and latitude values to each record. 
 We later selected only the books that had an entry for the country column. The original dataset contained Western numbers (e.g., 1, 2, 3); therefore, the native Arabic-speaking authors manually created a version with Eastern numbers 
% (\eg \<١، ٢، ٣>). 
\foreignlanguage{arabic}{١، ٢، ٣}.

%We used CSV files in our evaluation.
The dataset is available in the supplementary material. The aim of preparing this dataset was to create a \textbf{line chart} which displays the number of publications in each year, a \textbf{bar chart}, which displays the number of publications in each country, and a \textbf{map chart} displaying the number of publications in a selection of Arab countries. An example of these charts is shown in \cref{fig:teaser} (all charts are in supplementary material). 

% \begin{figure}[t]
%     \centering
%     \includegraphics[width=0.9\linewidth]{figures/Default-charts-Arabic-num.pdf}
%     \caption{Bar, line, and map charts generated using default settings in popular visualisation tools with Arabic text and Eastern numbers. 
%     \label{fig:default-charts}
% \end{figure}

%https://medium.com/@amnahhmohammed/useful-arabic-datasets-for-machine-learning-engineers-working-in-nlp-d06ba6c5e96d 

\subsection{Coding Process}
We did not find existing evaluation criteria of visualisation authoring tools in terms of language and direction support; our coding was informed by prior work evaluating visualisation authoring tools generally. Using a deductive approach, we annotated each tool based on three categories: tool characteristics, community support, and default charts. The tool characteristic category was drawn from related work (\cite{Huang2022AResearch, behrisch2018commercial}), which included chart type support, interactivity, and tool usability. Next, the community support category was informed by the compatibility support criteria of Huang \etals for visualisation authoring tools regarding language and direction support\cite{Huang2022AResearch}. 
Finally, the default chart category was drawn from Behrisch \etals \cite{behrisch2018commercial} and Alebri \etals work \cite{alebri2024design}, where the data type support (e.g., text and numerical rendering) and the visual element's positions and directions in a chart were evaluated. % We used a deductive approach and annotated each tool based on its characteristics, support and the default visualisation outcome, similar to prior work \cite{alebri2024design, Huang2022AResearch, lousa2019evaluation, behrisch2018commercial}. 
The first and third authors, who carried out the coding process, are native Arabic speakers, which is important for exploring the tools in Arabic. 
Each of these coders carried out the coding process independently and met afterwards to resolve disagreements by either recreating the default chart or testing a feature. 
%The coding process started by generating the most common basic charts (bar, line, and maps) as Alebri et al. \cite{alebri2024design} demonstrated. 
The visualisations (bar, line, map) were generated using the default settings, without any modifications, to ensure consistency in our analysis. Each coder captured a screenshot of the output for comparison and conducted the coding during the same period to minimise discrepancies caused by potential tool updates.

\subsection{Evaluation Criteria}
%We used a deductive approach and classified our codes into three categories inspired by prior work \cite{}.
We considered the tool's characteristics, community support, and the default chart type (bar, line, map). See \cref{tab:tool-codebook} for a detailed codebook definition. 
First,\hlthistleone{tool characteristics} included the following elements: \textit{Tool support for Arabic language, UI support for RTL, platform access, license, primary user group, chart types supported, RTL mirroring support, interactivity, calendar format supported, and font support}. Second,\hlLightOrange{community support} consisted of \textit{Arabic documentation availability and community support for RTL issues}.
Third,\hldarkseagreen{default chart} coded the directionality, position, textual, and numerical rendering of bar, line, and map charts generated using the default setting. 
For maps, we restricted our evaluation to choropleth maps as they are widely supported by most visualisation tools. %The classification included: \textit{the accuracy of Arabic text rendering, Bi-directional text handling, RTL mirroring support, Eastern number rendering, legend and label alignment, Axis based charts (X-axis direction and Y-axis position), and tooltip and annotation position}. 
The default chart evaluation was performed twice for each tool, using Eastern and Western numerals.

\subsection{Results and Analysis}
We assessed the tools’ characteristics and capabilities for supporting Arabic scripts by rendering our Arabic datasets to each tool to evaluate their default configurations. \textit{Overall, we observed a lack of support for Arabic scripts, particularly for Eastern numerals as shown in \cref{tab:default-chart-tool-analysis}}. Microsoft Excel performed best among the tools, particularly for its characteristics (see \cref{tab:tool-analysis}), and for its consistency in generating default charts for both datasets. Due to space constraints, we only report the most important results in \cref{tab:tool-analysis} and \cref{tab:default-chart-tool-analysis}.
We also assign binary values to features. For instance, for the \hlthistleone{RTL mirroring support} feature, we consider a tool to possess this feature if it provides any form of mirroring support (\eg text, only X-axis or offer it only on the browser version). For full details, refer to the supplementary material. Next, we report our key results based on our evaluation criteria.
%overall: lack of consistency? 

\subsubsection{Tool characteristics and community support}

The majority of the tools (above 71\%) offer some kind of \hlthistleone{RTL mirroring support} to reverse any elements of the chart (see \cref{tab:tool-analysis}). For instance, Flourish allowed customising the reading direction for text elements and icons only. Google Sheets allowed reversing the X-axis only. Power BI allowed mirroring only the Y-axis position. On the other hand, Excel and Datawrapper allowed reversing both axes. We also observed an inconsistency in the label for the RTL mirroring feature. For instance, Excel labelled it as ``Categories in reverse order'', Datawrapper required specifying the output locale under layout, Flourish used ``Reading direction'', Power BI used ``switch axis position'', and Google Sheets used ``Reverse axis order''. 

%The least-supported feature across tools is \hlLightOrange{custom plugin support} as shown in \cref{tab:default-chart-tool-analysis}, which can be beneficial for adding RTL functionality. Second, 
\hlthistleone{Calendar format} was poorly supported by most tools where alternative calendars to the Gregorian calendar were not offered. On the other hand, Microsoft Excel offered Hijri and Um Al Qura, which are common in Arab countries. %Google Sheets supports Hijri calendar only by coding using Apps Script. 
Most tools (above 57\%) provided Arabic \hlthistleone{font support} and had active \hlLightOrange{community support for RTL issues}. We found that most of these tools offered one or two default Arabic fonts. Yet, we did not investigate the level of support.   

RAWGraphs showed the weakest support for Arabic content (see \cref{tab:tool-analysis}), which is unsurprising given its origin as an open-source academic project built for research rather than commercial deployment. Surprisingly, mainstream tools such as Tableau, Flourish, and Datawrapper did not perform substantially better, indicating that even mature industry products have not prioritised Arabic language integration. %Tableau provide the tool in languages such as Hindi \& Japanese  
\definecolor{squareblue}{HTML}{666666}
\definecolor{squaregray}{HTML}{EEEEEE}
\definecolor{squarelightblue}{HTML}{6792F1}
\newcommand{\squaresize}{11.5pt}
\newcommand{\smallsquaresize}{8pt}

\newcommand{\lbsq}{\raisebox{-0.3ex}{\textcolor{squarelightblue}{\rule{\squaresize}{\squaresize}}}}
\newcommand{\gsq}{\raisebox{-0.3ex}{\textcolor{squaregray}{\rule{\squaresize}{\squaresize}}}\xspace}
\newcommand{\wsq}{\raisebox{-0.3ex}{\textcolor{white}{\rule{7pt}{7pt}}}}

\newcommand{\bsq}{\raisebox{-0.3ex}{\textcolor{LightPurple}{\rule{\squaresize}{\squaresize}}}}
\newcommand{\smallbsq}{\raisebox{0ex}{\textcolor{LightPurple}{\rule{\smallsquaresize}{\smallsquaresize}}}\xspace}

\newcommand{\ysq}{\raisebox{-0.3ex}{\textcolor{LightOrange}{\rule{\squaresize}{\squaresize}}}}
\newcommand{\smallysq}{\raisebox{0ex}{\textcolor{LightOrange}{\rule{\smallsquaresize}{\smallsquaresize}}}\xspace}
\newcommand{\smallgsq}{\raisebox{0ex}{\textcolor{squaregray}{\rule{\smallsquaresize}{\smallsquaresize}}}\xspace}

\newcommand{\hpad}{3pt}
\newcommand{\rotangle}{90}

\begin{table}[h]
\centering
\setlength{\tabcolsep}{\hpad}   % remove horizontal padding
\renewcommand{\arraystretch}{1} % keep vertical spacing normal

\caption{The analysed visualisation authoring tools and their support for Arabic. Purple \smallbsq indicate the availability of the tool characteristic, blue \smallysq denote the availability of community support, and \smallgsq refers to no support. The full analysis in the supplementary material. }%, while Grey squares \gsq\ represent the unavailability of the support.  }
\label{tab:tool-analysis}

\begin{tabular}{@{}rccccccc@{}}

& \rotatebox{\rotangle}{MS Excel} 
& \rotatebox{\rotangle}{Tableau} 
& \rotatebox{\rotangle}{Power BI} 
& \rotatebox{\rotangle}{Flourish} 
& \rotatebox{\rotangle}{Datawrapper} 
& \rotatebox{\rotangle}{Google Sheets} 
& \rotatebox{\rotangle}{RAWGraphs}  \\

%& \multicolumn{7}{c}{\cellcolor[HTML]{6792F1} Tool characteristics \lbsq} \\

{Tool Arabic Support} & \bsq & \gsq & \bsq & \gsq & \gsq & \bsq & \gsq  \\

{UI support for RTL} & \bsq & \gsq & \bsq & \gsq & \gsq & \bsq & \gsq  \\

{RTL mirroring support} & \bsq & \gsq & \bsq & \bsq & \bsq & \bsq & \gsq \\

{Calendar Format} & \bsq & \gsq & \gsq & \gsq & \gsq & \gsq & \gsq \\

{Font Support} & \bsq & \bsq & \bsq & \gsq & \gsq & \bsq & \gsq \\

%{Arabic Documentation} & \bsq & \gsq & \bsq & \gsq & \gsq & \bsq & \gsq \\ 

%{Community Support} & \bsq & \bsq & \bsq & \gsq & \gsq & \bsq & \gsq \\ 

%& \multicolumn{7}{c}{\cellcolor[HTML]{6792F1} Community support \lbsq} \\

{Arabic documentation} & \ysq & \gsq & \ysq & \gsq & \gsq & \ysq & \gsq  \\

{Support for RTL issues} & \ysq & \ysq & \ysq & \gsq & \gsq & \ysq & \gsq  \\

%{Custom plugin support} & \gsq & \gsq & \gsq & \gsq & \gsq & \gsq & \gsq \\

% \multicolumn{8}{l}{} \\
% \cline{1-1} \\[-8pt]
% \multicolumn{8}{l}{\smallbsq indicates the availability of the tool characteristic. } \\
% \multicolumn{8}{l}{\smallysq denote the availability of community support}

\end{tabular}
\end{table}

\subsubsection{Default charts}
\label{sec:default-charts}
We evaluated the \hldarkseagreen{default charts} generated by the seven visualisation tools using the same dataset. 
We considered the tool successful, unless it rendered a chart where the numbers treated as text or converted them from Eastern to Western. Overall, as \cref{tab:default-chart-tool-analysis} shows, Microsoft Excel most consistently supported Arabic for generating the three chart types, regardless of the numeral type. None of the tools generated a mirrored chart (i.e., RTL X-axis and Right Y-axis) by default. All the axis-based charts we rendered with default settings applied the LTR convention (\eg RTL X-axis and Left-aligned Y-axis). %Table \ref{tab:default-chart-tool-analysis} provides an overview of the tool support in generating Arabic visualisation using the default settings. 
Next, we report our findings based on the \textit{chart type, numerical handling, and textual support}. 
% We also display the results visually in \cref{fig:default-charts}.   

\input{default-chart-table}
\paragraph{Chart type}

Bar and line charts were generally well supported with Western numerals. Three tools failed to render maps properly, with RAWGraphs lacking map support altogether. In contrast, charts using Eastern numerals were poorly supported. Only Microsoft Excel successfully rendered bar, line, and map charts using Arabic text and Eastern numbers. Several tools were not able to render the charts properly. For instance, Tableau and Power BI treated numbers as text (see \cref{fig:teaser}). In some cases (\eg Flourish and Datawrapper), the output included axes but lacked marks and channels; in others (\eg RAWGraphs and Datawrapper), no output was generated. Additionally, some tools (\eg, Google Sheets) rendered the charts by automatically converting the Eastern numbers to Western, indicating a lack of native support for \hldarkseagreen{Eastern numeral rendering}. 
The rendering of visualisations depended on the chart type and the numeral system (see \cref{tab:default-chart-tool-analysis}). 
%Bar and line charts were largely stable with Western numerals but frequently failed or rendered blank with Eastern numerals.%, while 
Regarding maps, they were consistently the least reliable across both numeral systems.

%When using a dataset with Western numbers, all seven tools successfully generated a bar chart and a line chart, while only Microsoft Excel, Power BI, and Google Sheets created the map chart. With Arabic numerals, results were weaker. The map chart was the best supported, since it does not depend much on the text direction: Microsoft Excel, Power BI, and Google Sheets were able to display it, although Power BI showed Arabic numerals as text. In contrast, Flourish, Datawrapper, and RAWGraphs failed to create the map chart at all when using the dataset with Eastern numbers.

\paragraph{Numerical handling}
\label{sec:number-handeling}
\hldarkseagreen{Eastern numerals} were poorly supported in most  tools as shown in \cref{tab:default-chart-tool-analysis}. %(Tableau, Power BI, Flourish, Datawrapper and RAWGraphs) where they were not recognised automatically as numbers. 
For instance, Power BI misrecognised the data type for the bar chart. Manual attempts to change the data type (\eg in Tableau) were rejected by the tool, indicating limited flexibility in handling Arabic inputs. On the other hand, we successfully rendered all chart types in the seven tools when using Western numerals, except in three cases, due to text handling for maps in Flourish and Datawrapper, and because maps are not supported in RAWGraphs. Furthermore, we observed that Google Sheets automatically converted Eastern numerals to Western numerals without notifying the user. Although this conversion enabled chart creation, it reduces user control and raises concerns about the visualisation's integrity.

%\paragraph{Tool-specific observations}
%Only Microsoft Excel and Google Sheets consistently worked with both datasets across all three chart types, making them the most reliable options for RTL visualisation. Power BI provided partial support, producing the map chart using Eastern numbers but treating numerals as text. Flourish, Datawrapper, and RAWGraphs performed the worst overall, especially when using Arabic numerals, where they were unable to display some charts entirely. RawGraphs also showed problems when handling Arabic script, such as failing to display country names written in Arabic.

\paragraph{Text support}
Overall, the seven visualisation tools had better \hldarkseagreen{Arabic text rendering} than Eastern numerals rendering. Labels, headers, and other Arabic rendered correctly in most cases. 
Datawrapper and Flourish were the only tools that did not render Arabic text properly in maps, either causing an error or returned null values. Given the results presented above and the findings on text handling, there is a systematic mismatch between the two processes, resulting in a mixed presentation where Arabic text coexisted with Western numerals—highlighting the tools’ partial and inconsistent localisation support. 

Our analysis revealed that generating maps with default settings produced inconsistent region names and boundaries. 
% This suggests an embodied, hidden geopolitical bias. 
While this reflects ``prevailing geopolitical realities'', it also suggests that ``different perspectives are not equally embedded in tool defaults''.
For instance, Tableau and Google Sheets recognised the country entry ``Palestine'' whereas Datawrapper and Microsoft Excel converted it to ``West Bank and Gaza'' without notification. %We also observed a case where ``Palestine'' was treated as a null value. 
A similar inconsistency emerged with the representation of Western Sahara: while Microsoft Excel considered it as part of Morocco, others (\eg Tableau, Google Sheets, Datawrapper, Flourish) displayed it as an independent territory. %This inconsistency highlights the absence of standardized geographic data handling across tools and reflects underlying political and data-source dependencies in map rendering.

%Findings reveal that while Microsoft Excel demonstrates strong native support for Arabic text, numerals, and RTL layouts, other tools such as Power BI, Google Sheets, Flourish, Datawrapper and RAWGraphs offer partial or limited support. Persistent challenges include improper rendering of Arabic numerals, absence of localized documentation, and limited community discussion on RTL issues. Table 2 summarizes these findings.

%The results for testing the visualisation tools showed that some tools, such as Microsoft Excel, have strong support for Arabic, while others still have many problems, especially with Eastern numbers and layout direction. Many tools do not handle RTL text correctly or do not offer Arabic language help and options. This shows a clear need for better design and support for RTL users. Helping these tools work better with Arabic and RTL scripts will make them more useful and fair for many people around the world. 

%In contrast, tools such as RAWGraphs and Flourish showed broken Arabic letters or text that were not properly aligned when using Arabic script. This was particularly evident in cases where Arabic-Indic (Eastern) numerals failed to display at all or were treated as Left-To-Right text. It was also observed that in tools like Tableau and Google Sheets, while Arabic text was displayed, there was no native or built-in RTL chart orientation, and manual adjustments were required. This shows that RTL compatibility remains inconsistently implemented across the mentioned tools.

% END TEXT BY YASSINE
\section{Interview Study}
To further understand how the limitations of visualisation authoring tools impact the design process, we ran a follow-up qualitative study. 
We also explore design strategies specific to Arabic scripts, along with the methods that visualisation creators use to address these limitations. 
% Next, we outline our methodology. 

\subsection{Method}
\label{sec:interview-method}
%- ethics board approval and compensation \\
%- recruitment criteria- go back to the survey - whom we excluded \\
%- recruitment period \\
Our study was approved by the United Arab Emirates University's Social Sciences Ethics Committee (ERSC-2024-5388). We obtained informed consent from all interviewees, who filled in and signed a consent form (see supplementary material).
The interview addresses Alebri \etal \cite{alebri2024design}'s open research question about how designers reason about RTL adaptations in Arabic data-driven articles.
Accordingly, we expand our recruitment to visualisation practitioners who describe their work across design, journalism, and data analysis (see \cref{table:participants-demographics}).

We recruited our participants between December 2024 and May 2025. 
Potential participants completed a screening survey (offered in English and Arabic) and, if eligible, were contacted to schedule an interview via Calendly. 
Participants were eligible if they were \textit{native Arabic speakers} and \textit{designed Arabic data visualisations} as part of their work. 
We did not specify the level of expertise required to create visualisations. 
%how we recruited participants? how many institutions we approached? how many individuals we approached? what channels we used (e.g., Linkedin, DVS) \\
Our participants were recruited via a study poster on social media platforms (LinkedIn, X), mailing lists (ArabHCI, Data Visualisation Society), and snowball sampling. We also contacted the journalists who produced the visualisations included in Alebri \etal's \cite{alebri2024design} dataset, %(e.g., ARIJ, Inkyfada, Aljazeera). 
other journalism agencies in the Arab world, along with academic and research institutions,
%, including Arab Facts Hub and the Middle East News Agency. 
%Academic and research institutions were also approached.
%For example, we communicated with the research team from the Arab Barometer project at Princeton University, Qatar Computing Research Institute, and various universities in the UAE, Jordan, and the UK. 
commercial organisations that provide visualisation services to Arab clients (\eg Prezlab, Datawrapper), and journalists and data visualisation designers from our professional networks. 
Of 72 potential participants, 24 signed up for our screening survey; however, six of these were ineligible. 
We invited the rest ($n = 18$) for an interview, 11 of whom responded and attended the interview. Recruitment was conducted alongside data analysis, with the study being advertised while interview transcripts were being coded.

%- interview language \\
%- via Teams \\
%- The interviewer's language \\
%- interview length (around 35 minutes)\\
%- interview protocol (the questions asked in the interview)
Interviews were conducted on Microsoft Teams by a native Arabic speaker fluent in English. 
Participants could choose between Arabic and English to express their experiences without a language barrier. Eight participants chose to interview in Arabic.
% The interviewer was a native Arabic speaker and fluent in English.  
After introducing the participant to the purpose of the interview and their rights, participants discussed: (a) their background, (b) visualisation guidelines and design strategies followed, and (c) visualisation tools used. Participants were invited to walk the interviewer through a visualisation example they designed or share samples of their data-driven articles. 
% Participants were compensated for their time with an Amazon gift voucher or equivalent, worth £20. 
The average interview length was 35 minutes, and participants were compensated with an Amazon gift voucher or equivalent, worth £20. 
Our participants resided in five Arab countries, all reported having lived in one for at least 10 years, and nine reported fluency in at least one additional language. \Cref{table:participants-demographics} shows an overview of the participants. 
As demonstrated, there is overlap between the visualisation tools reported by participants and those analysed in \cref{sec:tool-analysis}. 

\begin{table*}[th]
  \caption{%
  	Overview of our interview participants' demographic data
  }
 \label{table:participants-demographics}
  %\scriptsize%
  \centering%
  \begin{tabu}{%
  	  r%
  	  	*{7}{c}%
  	  	*{2}{r}%
  	}
  	\toprule
  	P\# & {Profession} &   {Visualisation Tools/ Frameworks Used} &{Years of Experience} &  {Age Range} & {Sex} \\
  	\midrule
  	P1 & Journalist   & Flourish, Datawrapper, Power BI, Tableau &+16 & 35-44 & M  \\
  	P2 & Journalist     &Flourish, Datawrapper & 8-15 & 35-44 & M \\
  	P3 & Journalist &Flourish, Datawrapper, Excel &  1 & 35-44 & M \\
  	P4 & Data Analyst    &Flourish, Datawrapper, Power BI, Tableau & 1-3 & 25-34 &   M\\
  	P5 & Journalist &Flourish, Datawrapper, Tableau &  8-15 & 25-34 & F\\
  	P6 & Journalist   &Flourish, Infogram &  8-15 & 35-44 & M \\
  	P7 & Journalist &Flourish, Datawrapper, Python & 8-15  & 35-44 & M \\
  	P8 & Designer     & Datawrapper, Excel, RAWGraphs   &  4-7 & 25-34 & M \\
  	P9 & Designer     &Flourish, Datawrapper, Cables, D3, R &  8-15 & 35-44 & M \\
  	P10 & Journalist  &Tableau, Excel, R &    +16 & 45-54 &  M  \\
  	P11 & Journalist \& Student    &Canva & 1-3  & 25-34 & F \\
\bottomrule
  \end{tabu}%
\end{table*}

The first, fifth, and sixth authors (native Arabic speakers) edited the Arabic interview transcripts, and the second author edited the English ones. Editing the Arabic interviews required first standardising the colloquial dialect to Modern Standard Arabic to ensure consistency across the Arabic dialects and maximise the quality of the English translations by Google Translate \cite{sabtan2021evaluation}. 
% Ensuring that all transcripts are in Modern Standard Arabic maintains consistency, as Arabs coming from 22 countries speak different dialects depending on their location. %The Arabic transcriptions were then translated to English using Google Translate, and
The first author then revised these translations against the original to ensure their validity. All transcripts were anonymised before translation and their flow enhanced with any visualisations that the participant discussed. % The identity of the participant was removed from all transcripts before translation. 

The transcripts were analysed using thematic analysis \cite{Braun2006UsingPsychology}. 
The first and second authors began with independent coding and developed their own codes and interpretations.
After completing every one or two transcripts, they met to discuss their codes. % into theme sets. 
Overlapping codes were renamed when needed and merged. 
For conflicting assignments, the coders discussed their rationale and interpretations in relation to the research questions: consensus resulted in applying one or both codes, or a new code combining both ideas, and disagreements were tracked. 
We held regular group meetings with the other authors to discuss codes, update the codebook, and resolve disagreements. 
We decided to stop recruitment when the transcripts showed significant repetition and reached saturation.
The group's meetings also led to identifying and refining the themes to those supporting our research thesis.
For instance, a theme on the field of data journalism was discarded: though recurring across participants, it mainly covered codes such as workflow that reflected high-level organisational structure without a clear link to design decisions.
Whenever new codes emerged, we revised the coding process for the previous transcripts. 
Our process yielded 26 codes clustered into four final themes (see Supplementary Material).

\subsection{Results} 
%limitation circle 
\newcommand{\circledL}[1]{%
\tikz[baseline=(char.base)]{
\node[
    circle,
    draw,
    fill=lightpeach,
    inner sep=0.6pt,
    minimum size=1.4em,
    font=\bfseries\small,
    text height=1.4ex,
    text depth=.25ex
] (char) {L#1};
}%
}

%Adressing circle
\newcommand{\circledA}[1]{%
\tikz[baseline=(char.base)]{
\node[
    circle,
    draw,
    fill=lightbluecustom,
    inner sep=0.6pt,
    minimum size=1.4em,
    font=\bfseries\small,
    text height=1.4ex,
    text depth=.25ex
] (char) {A#1};
}%
}
%RQ:How do visualisation creators navigate design decisions when creating RTL visualisations, particularly for Arabic scripts? What tool features or guidelines can better support them in addressing script-specific and cultural challenges during the design process?

We identified four themes to explore practitioners' visualisation literacy, design strategies, challenges, and tools used during the design process for Arabic visualisations.  %:\textit{The designer's identity, systems of power, imagined audience, design strategies, and visualisation tool limitation}. 
%Next, we explore how Arabic visualisation creators navigate design decisions, the challenges they observe, and the support they need. 

% We used thematic analysis \cite{Braun2021ThematicGuide} to analyse the interview transcripts. 
% Four themes emerged that evolved around 
% how designers design within the constraints of tool support for Arabic scripts (\cref{theme:design-strategy-limitation}), 
% how designers' visualization literacy (\cref{theme:designers-literacy}) and assumptions about the audiences' (\cref{theme:designers-assumptions} shape their design decisions, and 
% the power emebedded in visualisation tools' defaults (\cref{theme:interview-power}).

% transcripts to establish our codebook.

%%%%%Themes%%%%%%
\subsubsection{Designing Within Constraints}\label{theme:design-strategy-limitation}
%Addressing Tool Limitations for Arabic Scripts}}

This theme presents the key \hlred{tool limitations} shared by participants, followed by the strategies they employed to \hlblue{address} these challenges in practice. 
%closing sentence: Together, these accounts reveal how designers negotiate the boundaries of tool capability through situated acts of adaptation.
%lack of mirroring features (P5, P8) 
Participants explained that the tools lacked features that enable 
\hlred{automatic mirroring}. Mirroring is a design strategy used to flip the direction and position of UI elements based on the reading direction. %(see \cref{fig:teaser} bottom center). 
As prior work suggests, this is a common strategy found in Arabic visualisations \cite{alebri2024design}. 
%P5 explained \aquote{``when you write this chart in Arabic, you can't really flip it''.} 
P8 reflected on their experience with Excel and explained:
\begin{comment}
   \begin{bluequote}
not everything is flipped actually. It can read the data ...
the data ... %I mean the text entry, 
but it doesn't flip. The legend at the top stays as is, so it stays from left to right...that is time-consuming because I have to do it manually. I have to download it as an editable version where I lose the feature of interactivity
\end{bluequote} 
\end{comment}
\aquote{``not everything is flipped actually. It can read the data ... %I mean the text entry, 
but it doesn't flip. The legend at the top stays as is, so it stays from LTR...that is time-consuming because I have to do it manually. I have to download it as an editable version where I lose the feature of interactivity''.}
% {P8}
In parallel, P5-6 also found the manual labour to edit the visualisation due to the lack of the mirroring feature to be time-consuming and frustrating, as P6, who is a freelancer and manually mirrors chart components using a photo editor, stated \aquote{``it's a very frustrating process, very time-consuming. Another journalist can build this in 10 minutes... But for me it would take more than a couple of hours''(P6).}
% {P6} 
P2, P5, and P8 expressed willingness to use the mirroring feature if integrated into visualisation tools.

    %tension between Alignment of language and visual interpreation (P2-3, P5-6, P9) and Seperation of language and visualisation interpretation (P1, P8)
Our analysis revealed a \textit{tension between aligning the language and visual interpretation and separating the two}. While some participants aimed to preserve the Arabic reading flow by aligning text and chart direction (P2-3, P5-6, P9), others deliberately separated the two (P1, P4, P8) to maintain visual clarity and avoid distortion. On the one hand, some participants saw that the entire chart needs to align with the Arabic script. For instance, P9 stated that it is not enough to simply translate the text and stop there without mirroring and justified this position by stating \aquote{``It is normal when I open an Arabic book, I open it from the right''}. %, but I open an English book from the left''}. 
Similarly, P2 expressed that they would choose mirroring a chart for the following reason \aquote{``If my story was written in Arabic, why don’t I make the chart more Arabic, as we learned, so that it starts from RTL?''}. % I don’t write in a foreign language''}. 
Also, P3 who chooses only to translate the text in a chart stated \aquote{``I think it's best to display it from RTL since we're writing the axes in Arabic. It's better to be consistent''}. 

On the other hand, some participants saw that the language direction should be viewed in isolation from the chart direction. For instance, we heard from P1 \aquote{``Some people tend to think that the direction of everything in Arabic is from the right, and this is actually confusing''} and commented that mirroring is only for the textual elements in the chart. Similarly, P8 expressed \aquote{``chart is a chart, right? you just need to translate the text, not the way that we show the information''}.   
 
    %mirroring process (P1, P4, P6, P8-9)
While the tools lacked automatic mirroring features, participants (P1, P4, P6, P8-9) expressed that they mirror their charts when designing for Arabic content and use the following strategies to address this shortcoming in visualisation tools. 
One of these strategies was \hlblue{using a second tool}, whether it was a photo editing tool or custom coding to modify the chart. Our participants reported using the following tools to edit their visualisations: Adobe Photoshop ($n = 3$), Canva ($n = 2$), Adobe Illustrator ($n = 2$), Adobe After Effects ($n = 1$), Figma ($n = 1$), and Microsoft PowerPoint ($n = 1$). %as shown in \cref{table:participants-demographics}. 
%P6, P8-9 shared (see \cref{table:participants-demographics}). 
% \Cref{table:participants-demographics} shows  what participants used Canva (n = 2), Adobe After Effects (n = 1), Adobe Photoshop (n = 3), Adobe Illustrator (n = 2), Figma (n = 1), and Microsoft PowerPoint (n = 1) to edit their visualisations. 
% \alongquote{``for every part of the infographic I had to photo edit it to mirror images to make the flow of information visually from RTL, and then start again putting labels in Arabic, align them to the right.''}{P6}    
\hlblue{Removing the axes} was another strategy participants practiced (P1, P4, P9) while mirroring visualisations. Participants highlighted that removing the axes entirely reduces confusion. %\textit{``I might remove the y-axis line entirely and leave the points written on the grid lines in the middle to avoid interfering with the graph's concept''} and 
% P1 shared \textit{``If the bars are directly labeled ...sometimes we get rid of this confusion by not having an axis at all''}. 
\aquote{``If the bars are directly labeled ...sometimes we get rid of this confusion by not having an axis at all''} as P1 stated. 

P4 also suggested a \hlblue{temporary guiding solution}: \aquote{``to highlight the axis to clarify its beginning''} until readers are used to the direction. 
    %Data type dictates orientation (P4, P6)
The absence of automated mirroring features pushed participants to resort to \hlblue{data-driven and context-sensitive reasoning}. For instance, P4 remarked \aquote{``I think that if the x-axis contains qualitative data and not quantitative data, then I think that there might not be a problem in reversing the axis because ...it doesn't logically conflict with the direction of the Cartesian axis''}. 
On the other hand, P6 stated that reversing shouldn't be implemented on all \textit{axis-based charts}, but can be safely applied to \textit{flowcharts}.
    
%Numbering challenge (P1, P4-6) (Done)
Participants expressed \hlred{challenges with Eastern numerals} when designing Arabic visualisations. 
P6 asserted this difficulty, which they addressed by \hlblue{opting for a simple or basic chart}.  %\aquote{``if I decided to use the [Eastern numbers] it was more difficult, of course, so I ended up most of the time just trying to have very basic charts''(P6).}
% {P6} 
P5 had a similar remark: \aquote{``It's either not recognising the [Eastern numbers] or it lets you write the numbers in the [Western] way... it kind of messes up the whole text you are trying to write...So this leaves you with one or two platforms that you can work with''.} 
Furthermore, P4 expressed great difficulty when adding Eastern or Western numerals to Arabic text (\eg, titles or annotations) as the order gets ``messy". 
% 
% 
%mixed language presentation (P10, P7)
Thus, participants were forced to use a \hlblue{mixed language presentation} in their produced visualisations. For instance, P10 expressed \aquote{``the statistical reports I produced for the company had numbers written in English, but the text itself was written in Arabic''}. 
Similarly, P7 stated \aquote{``in Arabic, this is how the visualisation looks [refers to a LTR scatter plot with Arabic textual annotation], and the numbers themselves are Western numbers''}.   

% text-related limitations
Participants described multiple \hlred{text-related limitations}, demonstrating a lack of infrastructural support for Arabic scripts. 
%text alignment issue 
The most common issue that participants (P4-7, P11) faced when designing Arabic visualisations was the \textit{text alignment problem}. 
% P11 shared: 
\aquote{``Even at the writing level, there is anxiety \ldots Sometimes even when you enter the address in Arabic, it basically puts the letters together and does not write the words as you want them (see \cref{fig:teaser})''(P11).}
% {P11} 
% (also see \cref{fig:teaser} for an example) 
P7 also stated that Python creates Arabic visualisations with disconnected letters (see \cref{fig:teaser}). 
Yet, even some mainstream visualisation tools do not support Arabic script properly. 
%The size of the company that developed the visualisation tool is not reflected in supporting a basic need for RTL scripts, as P4 remarked 
\aquote{``A platform like Tableau, with its size and presence in the Arab world, still has alignment issues in Arabic text \ldots especially when you put numbers in the middle of the text (P4)''.}
% {P4}  
\begin{comment}
   \begin{figure}[h]
    \centering
    \begin{minipage}{0.48\linewidth}
        \centering
        \includegraphics[width=\linewidth]{figures/Fixed-Arabic-bar.png}
        \caption{An Arabic bar chart that displays text properly.}
        \label{fig:example-good-chart}
    \end{minipage}
    \hfill
    \begin{minipage}{0.48\linewidth}
        \centering
        \includegraphics[width=\linewidth]{figures/messy-Arabic-bar.png}
        \caption{An Arabic bar chart with separate letters, making the chart unreadable.}
        \label{fig:example-bad-chart}
    \end{minipage}
\end{figure} 
\end{comment}

% Lack of language support (interface & data entry) in tool is a barrier (P1, P4, P6, P7, P11)
Beyond text rendering issues, participants were challenged because of the \textit{absence of Arabic interface options and unreliable data entry for Arabic text}. 
For instance, P7 reflected on their interaction with Power BI and stated: \aquote{``the app's interface is completely unhelpful to Arabic readers. The interface requires knowledge of English or languages that use Latin script, and Arabic isn't supported''}. 
Similarly, P8 noted that these tools do not automatically recognise Arabic: \aquote{``in most of the programs that I use, they do not read Arabic until you do it manually''}. 
P11 also suggested that data entry is not as simple as it is for Latin languages: \aquote{``these tools are, even at the logistical level are always in French or English. It is true that someone who designs a chart in Arabic faces some difficulties even at the level of data entry.''}%{P11} 

%Lack of font support (P4, P9)
Furthermore, participants also described \textit{poor font rendering}. 
For example, P9 explained the absence of control with Arabic fonts: \aquote{``there aren't options to change fonts, there might be only one Arabic font available''}. 
P4 also explained that the nature of the Arabic language is not considered with 
the font options available in these tools: \aquote{``most Arabic fonts, when we 
try to add diacritics, 
% such as shadda (\<ّ>) and damma (\<ُ>), 
such as shadda \foreignlanguage{arabic}{ ّ} and damma \foreignlanguage{arabic}{ـُ}.
for example, the letters break (see \cref{fig:teaser}). There aren't many fonts, or many open fonts available, except for one or two, as far as I know, that have good design by Arabic standards''}.

% Addressing tool limitation (P1, P5-9)
In response to these issues, participants (P1, P5–P9, P11) responded by {\hlblue{opting for a simple or basic chart}} or avoiding certain charts. 
% For example, participant P5 commented 
\aquote{``in Arabic, the features that you can control are pretty limited. So, when I design a visualisation in Arabic, I usually tend to keep it nice and as simple as possible (P5)''.}
% {P5} 
Similarly, participants rely on other strategies such as \hlblue{producing the chart in a Latin language}:
\aquote{``most organisations in Tunisia, even large organisations which are affiliated with the state ... use data in French and produce visual presentations in French because it takes a lot of time when writing even a title in Arabic (P11).''} %{P11} 
P7 even shared that they end up producing the chart in simple English to avoid the difficulty of producing Arabic charts in Python. 

\subsubsection{Practitioners' visualisation literacies and design decisions}\label{theme:designers-literacy}

Participants positioned themselves in relation to different forms of literacy, such as graphical and English language literacy and to their awareness of tool mirroring features that shaped their design practices. 
Participants were divided on mirroring visualisations for Arabic scripts because of their \textit{graphical literacy} (P1, P4, P6, P8-9). 
Their reflections on the topic highlight the influence of their educational background.
% According to P4 who went to an Arabic school: 
\alongquote{``we did not reverse the axis from RTL for positive values \ldots Therefore, it has never happened in our culture in general or in our knowledge, to see the axes in reverse. Therefore, we have always relied on maintaining the direction of the chart as it is, from LTR.''}{P4}% -- went to an Arabic school}
\alongquote{``I mean, the X or Y axis starts from LTR based on maybe my education, I don't remember that we used to flip anything.''}{P8}
% On the other hand, P9 who went to an English school in Egypt commented 
\alongquote{``The Coordinate System, \ldots we learned it in school \ldots we got used to it this way \ldots The segment of society that studied Arabic \ldots Their direction was from RTL, the opposite of ours.''}{P9}% -- went to an English school in Egypt}  

The above quotations also suggest the importance of language literacy. While \textit{English language literacy} was not a major issue for most of our participants based on their educational background, some (P3, P7) expressed language difficulties. 
P7 took an Arabic Journalism Diploma where the teaching language was Arabic, described:  
% \textit{``the diploma helped me overcome the language barrier, as most of these tools are [in] English''}.
\aquote{``the diploma helped me overcome the language barrier, as most of these tools are [in] English''}.
P3 expressed concerns about not using the full capacity of visualisation authoring tools because of language barrier: 
% \textit{``there's a lack of understanding of English terminology. When we translate it into Arabic, we face difficulties. It might be a statistical term, and even if we understand the term after translation, it's difficult to understand the identity of the term, its meaning, or how to employ it. ...We may need more... explanations in Arabic about design... we're only using a small portion of what's available''}. 
\alongquote{``there's a lack of understanding of English terminology. When we translate it into Arabic, we face difficulties. It might be a statistical term, and even if we understand the term after translation, it's difficult to understand the identity of the term, its meaning, or how to employ it. ...We may need more... explanations in Arabic about design... we're only using a small portion of what's available.''}{P3} 
This strengthens the importance of efforts like The Data Visualisation Glossary \cite{ArabiDataVisGlossary} that translate technical terms in visualisation design into Arabic. 
Regarding awareness of mirroring features, P3 and P8 were not aware of them in some visualisation tools, although both of them use tools that provide those features (\eg Datawrapper, Excel)
\aquote{``I haven't looked into this feature, this is the first time I've heard of it''.}

\subsubsection{Practitioners’ assumptions about audiences}\label{theme:designers-assumptions}
%Graphical literacy assumption
Extending the previous theme, participants revealed how their assumptions about the audience shaped their designs, reinforcing universality. 
In particular, many participants (P1, P3-6) assumed a certain \textit{graphical literacy} about Arab audiences. For instance, P5 stated \aquote{``there are charts that I will not use in an Arabic article because I don't really believe that the Arab audience would understand''.} 
We also heard a generalisation of how Arab audiences read graphs from P3 \aquote{``I think most Arab readers are used to reading charts from LTR''.} 
Similarly, P6 opted to use simple charts because they assumed the Arab audience's graphical literacy \aquote{``I would say since most of the Arabic audience are not used to seeing this kind of data visualisation, we need to make sure that when we are using them, they have to be clear and self-explanatory''.} 
P1 also implicitly suggested an assumption about the audience's graphical literacy \aquote{``I personally think that changing the direction of the curves based on the language is a huge mistake and creates a state of misunderstanding among the audience who are used to reading curves and graphs in their precise statistical concept from the left.''}

%audience habituation assumption 
Participants also highlighted assumptions about the \textit{audience habituation} (P1-3, P9) to LTR charts in Arabic scripts. 
P2 stated \aquote{``because most platforms currently publish charts on social media, the reader's eye and the viewer's eye began to get used to the idea that the chart starts from LTR''}. 
P9, who designs mainly in English, made the following assumption \aquote{``I know that due to the lack of Arabic content specific to graphs, we are accustomed to English content, and thus we are accustomed to the direction from LTR''}.

%reading direction expectations
Furthermore, participants (P1, P6) also assumed the \textit{reading expectations} of the Arab audience. P6 generalised their personal experience to all Arab audiences \aquote{``even when I read something in English, by default my eyes go from RTL. When I look at a picture or any kind of visual I assume that this should be the case \ldots for our Arabic readers.''}
 P1, who stated views against mirroring, indirectly expressed a generalisation of what the audience's expectation are: \aquote{``Usually, there is a visual culture among people, and violating it is an obstacle.''}

%language literacy assumption (P6, P9)
Participants also made assumptions about \textit{the audiences' ability to understand the statistical language} used in their stories. 
For instance, P6 anticipated the following \aquote{``I feel like for one of the statistics [I] used, People are not familiar with it, I have to explain it''.} 
We also heard from P9 who justified this assumption by the fact that statistical terms are not used in colloquial Arabic \aquote{``when I was addressing the Arab audience, I was concerned not to give them a difficult word that they could not understand from the chart itself, such as a difficult definition or a difficult word used".}

\subsubsection{Embedded power in visualisation defaults}\label{theme:interview-power}

%English is THE language for learning -- disseminating in English is better
Participants' reflections revealed how the design decision process is restricted by political, economic, and epistemic control. Many participants (P1, P3-11) brought up issues with the design process, which hinted towards what we interpret as a \textit{colonial legacy} \cite{Dourish2012UbicompsImpulse}. In particular, this was in the shape of perceiving the content produced in Latin languages as superior (P3, P9-11). 
\alongquote{``I have to produce the rest of the projects in English due to its popularity, because projects that are only produced in Arabic do not receive the expected or anticipated response and engagement from the Arab world.''}{P9}
%\alongquote{``more knowledge is produced in foreign countries that use the left side''}{P3} 
\alongquote{``producing any content in English is more elegant,... we [Arabs] lack an aesthetic sense in producing statistical reports.''}{P10}

Yet, this perspective was also propagated in contemporary tool design, where \textit{English is treated as the default} while other languages are silenced (P4-8, P11). 
P8 asserted \aquote{``Of course it [data file] will be in English!''}. 
In another example, P5 reported their experience using Flourish, Datawrapper, and Tableau that limited their agency in making design decisions \aquote{``It [visualisation tool] messed up the design pretty much because the design template is designed for English speakers and language\ldots I avoid writing a complex text or complex headline''.}
Because English was the default, developing with it is a more convenient option as P10 elaborated  \aquote{``Any data I used was originally in English, and then I Arabized it. It's easier and faster to use English.''}

%Tool dictates direction (P2, P3, P6, P7)
The systems of power were not only visible in the privileging of English, but also in \textit{the ways tools enforce LTR defaults} (P2-3, P6-7), compelling participants to adapt their work to the tool’s assumptions. 
For instance, P2 shared their thinking process with the design team whenever they are borrowing an English visualisation: \aquote{``let’s stay on the same basis as Flourish itself, which forces you to start from LTR''}. 
P6 also made a generalisation of the visualisation tools by noting \aquote{``if you're using any tool, by default all the effects are from LTR''}. 
Beyond visualisation tools, P7 extended this concern to libraries \aquote{``Python, for example, does not ask about the direction. It assumes that the graph has the correct direction''} and ended up producing the chart in English.

%P3: \textit{``It's better from left to right, even for Microsoft Excel''}

%Normalization of LTR in Arabic conventions
The framing of English defaults as universal quietly reinforces LTR conventions in Arabic visualisations, revealing how technical settings can embody power.  
%For example, we heard from P2 
\aquote{``because most platforms currently publish charts on social media, the viewer's eye gets used to the idea that the chart starts from LTR. (P2)''}
% }{P2} 
% In parallel, P3 expressed 
\alongquote{``I think we've gotten used to the idea during our studies that charts are read from LTR, because it's more compatible and easier, even in Excel... It's easy to standardise, to the point that I think most Arab readers are used to reading charts from LTR.''}{P3} 
The lack of guidelines for Arabic visualisation design (P1, P3-10) reinforces the dominance of LTR conventions in Arabic scripts.
In some instances, our interviewees think that violating the normalised conventions would harm the reader: \aquote{``even if it is a mistake, sticking to it will facilitate the purpose of simplifying the conveyance of information to people and not confusing them (P1)''.}

% lack of localization for map design (P1, P6-7)
This normalisation extended beyond language and direction to spatial representation. Participants (P1, P6-7) expressed that local realities are replaced by treating the Western geopolitical framings as universal, just as visualisation tools treat English and LTR direction as universal. For instance, P1 and P6 expressed concerns about the misrepresentation of the Arab world, where certain areas are not even displayed on the map 
% (\eg Palestine) 
or enforced different labelling. 
P7 also highlighted the lack of recognition of country or city names in Arabic in these tools: \aquote{``Because the names of Arabic cities and governorates in Python are not available in Arabic, they must be entered in English for the map data to be understood.''} %While Datawrapper specialises in maps, it also requires me to translate each governorate into Arabic.''}{P7} 
This not only lengthens the design process but also prevents practitioners from depicting uncertainty in areas of conflict.

% Access issue (P4, P5)
Participants also discussed \textit{access issues}, noting that creating Arabic visualisations often required additional effort to mirror designs, while some necessary features were further restricted by paywalls: 
% For example P5 explained: 
\aquote{``most of these tools give you some control when you're a free user. But, they give you more control if you are a premium \ldots and the problem is that most of these platforms are not really affordable.''(P5)}

\section{Discussion}
    We combined analytical evaluation of visualisation tools with semi-structured interviews to understand how practitioners navigate design decisions when creating Arabic visualisations and how tools support them. Next, we map our results to relevant literature and outline a research agenda with recommendations for visualisation design practice. 
    % Our investigation combined analytical evaluation of visualisation tools with semi-structured interviews to understand how visualisation creators navigate design decisions when creating Arabic visualisations and how visualisation tools can support them. Next, we map our results to relevant literature and outline a research agenda along with recommendations for visualisation design practice. 

    %\subsection{Compromised creativity and Work intensity under low agency}
    % \subsection{Compromised creativity under low agency}
    \subsection{Constrained Work: Effort, Agency, and Creativity}
    \label{sec:working-under-constraints}
    %Low agency as designers often have no saying 
        %''Blackboxing functionality"
    %claim   
    Our results suggest that Arabic visualisation practitioners experience limited agency over the fundamental elements of their produced visualisations. 
    %evidence from findings 
    For instance, they feel compelled to use a basic chart, avoid specific charts, or even present it in English to work around the tool's limitations for RTL scripts (\cref{theme:design-strategy-limitation}). 
    Given that our analysed tool selection overlaps with the tools reported by participants, our tool analysis also confirmed this finding, in which we observed instances (see \cref{sec:number-handeling}) where the tool converted Eastern numerals to Western and forced a mixed presentation (see \cref{fig:teaser}) - a task that should be very basic in generating a visualisation in these tools.
    %link to previous literature
    Although such sense of low agency is not exclusive to Arab practitioners, as previous work suggests that automated visualisation tools often induce low authorship \cite{Mendez2017Bottom-upTools}, we demonstrate that this is even more pronounced when designing in Arabic, where the language itself is overridden. 

    %Fragmented design process leading to high effort on trivial matters 
    %claim 
    The low agency pushes the design process of Arabic data visualisation toward fragmented workflows, resulting in increased friction and effort over trivial issues (\eg, text alignment, the legend and the source position). 
     %evidence from findings 
    Participants reported that they often need to use multiple tools to achieve their desired final design (see \cref{theme:design-strategy-limitation}) or spend a considerable time translating each Arab city name (see \cref{theme:interview-power}). %This includes fixing trivial issues such as text alignment, the legend and the source position. 
    Using multiple tools may result in the loss of some capabilities, as our participants reported resorting to static visualisations because they must fix charts using other tools. 
    Although extending the tool's capabilities through add-ons or custom plugins could mitigate the problem, it does so at the cost of increasing the user's workload.
    %link to previous literature
    The time-consuming nature of this process places Arabic visualisation practitioners at a structural disadvantage in benefiting from the efficiency of mainstream visualisation tools as reported by Méndez \etal \cite{Mendez2017Bottom-upTools}.

    %Creativity is compromised because of the aforementioned difficulties 
    %Claim
    Due to the aforementioned constraints, creativity in Arabic data visualisation is compromised. 
    %evidence from findings 
    Participants reported frequently settling for simpler text and representations (\eg, basic charts) due to tool limitations, suggesting that the effort and time required to manage foundational elements—such as Eastern numerals and text alignment—discourages exploration of more complex or expressive designs. 
    %link to previous literature
    While Baigelenov \etal \cite{Baigelenov2025HowInspiration} has shown that practitioners often draw inspiration from presets and system recommendations, this inspiration can be limiting rather than generative, particularly when alternatives require substantial effort.
    Over time, the repeated circulation of such constrained outputs risks reinforcing ``design fixation'' \cite{DavidG.Jansson1991DesignFixation}- a cognitive barrier in which designers adhere to familiar solutions rather than exploring alternatives. 
    Our findings extend Parsons \etal's \cite{ParsonsFixationAnd} account of fixation attributed to presets, visualisation corpora, effort, and habitual practices by showing how these mechanisms are amplified in the context of Arabic-script visualisation.

    %implications 
    An implication of these findings is that visualisation tools offering automated design should critically revisit the level of control afforded to their users across scripts, ensuring a unified set of design elements that remain user-controllable and are not compromised by script direction, such as numeral systems or the set of recommended chart types. Tools can also be more transparent about the limitations imposed due to script language by notifying the user of any overridden changes. 
    Furthermore, visualisation tools should be designed to preserve continuous design flow by reducing interruptions not only for practitioners working with Arabic but for all practitioners, particularly at the level of foundational elements, and by supporting bottom-up creative processes rather than enforcing fixed layouts or script-dependent presets. %This leads to making the defaults in these tools to be flexible and less authoritative, especially for the foundational aspects of a visualisation. 

    \subsection{Exercises of Power in Visualisation Practice}
    \label{subsec:power-default}
   
    %superiority cycle 
    % claim #1
    Our findings show that power in Arabic data visualisation design operates through a structurally reinforcing chain that is difficult to interrupt, in which cultural beliefs, such as the perceived universality and dominance of English, become embedded in conventions, materialised through tool defaults, and normalised as standard practice.
    %evidence from findings
    As reported in \cref{theme:designers-assumptions}, some participants expressed concern that violating LTR conventions could cause confusion and disrupt message delivery. 
    Importantly, this concern appears to stem not only from perceptions of the reader's familiarity with LTR conventions but also from the absence of RTL conventions, thereby positioning deviation as a risk rather than a legitimate design choice.
    %link to previous literature
    While prior work critiqued conventions for the power they extend to visualisations, making them render as clinical and impartial \cite{Kennedy2016TheDo}, their absence is also problematic. 
    To illustrate, readers might think something is increasing when it is actually decreasing, simply due to the axis position or orientation. 
    Alebri \etal \cite{alebri2024design} found inconsistent design patterns when depicting numerical data in Arabic script, even within the same article, suggesting a lack of standards. This is similar to the case of depicting uncertainty, where practitioners may avoid representing uncertainty when guidance is lacking \cite{Hullman2020WhyUncertainty}. 
    Our work presents a case study of how cultural dominance shapes the visualisation practice through defaults. 
    %evidence from findings
    Participants repeatedly described how the design process in the visualisation tool, from uploading data to editing chart titles, was consistently governed by LTR defaults. This was further confirmed by our tool analysis, which showed that Eastern numerals were rendered as text, while Western numerals were treated as numerical values, reinforcing asymmetric support at a foundational level. 
     %link to previous literature
    Despite substantial attention to the rhetorical power of visualisations (\eg \cite{doerk2013CriticalVisualization, hullman2011visualization}), our findings highlight that, in the case of RTL scripts, this rhetorical power lies not with the authors but with the tool. We also demonstrate how visualisation authoring tools have political dimensions. %extend Costanza-Chock's work \cite{Costanza-Chock2019DesignPractice} by demonstrating how v. 
    Costanza-Chock argues \cite{Costanza-Chock2019DesignPractice} that defaults are shaped around an ``unmarked” imagined user defined by dominant cultural and linguistic norms, which are then internalised and reproduced through design systems rather than explicitly chosen by designers, demonstrating the artefact's power. %Even when writing this manuscript, we encountered challenges in including Arabic text within the template, suggesting a deviation from the norm. 
    Defaults matter particularly in visualisation, as the field serves users with diverse backgrounds and levels of expertise. 
    Moreover, defaults are often trusted, especially by novices who may feel unqualified to question them and accept them as given \cite{Lauer2020HowGraphs}. %Defaults also function as a source of inspiration in visualisation practice \cite{Baigelenov2025HowInspiration}, amplifying their influence on what designs are considered possible or appropriate.
    %Visualisations generated using defaults are often used when evaluating LLMs (e.g., ), therefore part of our knowledge of LLM capabilities are shaped by these defaults as the training of these tools is based on these default visualisations. 
    %Defaults matter as they contribute to our knowledge. 
    %For example, studies that evaluate LLMs (e.g., ) through default risk research contributions that mirror unintentional default biases more than genuine capabilities.

    % Maps and the worlds they hold 
    % claim #3
    A single worldview is embedded in most visualisation authoring tools when depicting maps of the Arab world, demonstrating a power structure. Country borders and their names change over time, and the current settings of these tools implicitly assume stability and reject alternative framings or ongoing conflicts.
    %evidence from findings
    Our interview participants commented that the tools misrepresent the Arab World by not recognising Arabic names and enforcing labels without notification (see \cref{theme:interview-power}). Our tool analysis affirms this, showing that Arabic country names were not even recognised by the tool (\eg Flourish). 
    %link to previous literature
    As Monmonier \cite{Monmonier1991HowMaps} and Winner \cite{Winner2017DoPolitics} demonstrate, map design elements such as heavy lines and area names contribute to sovereignty rather than a scientific representation of geographic reality. Although some of the visualisation tools allow users to upload their own map, non-experts with limited time may opt for the default map, leading to further exercise of that perspective. Displaying maps that omit or re-label some countries may lead readers to distrust or feel detached \cite{Koesten2025EncounteringMaps}.

    %authors hold power over the receivers of visualisations
    % claim #4
    While practitioners are constrained by external sources of power, such as tool defaults, our results indicate that they simultaneously mediate power over readers through assumption-driven visualisation design choices. 
    %evidence from findings
    Section \ref{theme:designers-assumptions} demonstrates how our interview participants avoided a certain visualisation or selected a particular language based on their assumptions about the reader. 
    %link to previous literature
    This finding further confirms the work of Hullman and Diakopoulos \cite{hullman2011visualization}, who argue that editorial design choices (\eg visual representation, annotation, omission, emphasis) are made and that visualisations therefore hold rhetorical power over certain interpretations. Readers are treated as passive recipients through these assumptions, amplifying the ``banking model'' critiqued by Freire \cite{Freire2005PedagogyOppressed}, in which he argues that knowledge should be co-constructed. 
    % rather than poured into them.  

        %implications

    The influence of power in the form of defaults extends beyond individual design outcomes: visualisations generated using defaults are frequently employed in evaluations of Large Language Models (LLMs) (\eg \cite{Rauf2025EvaluatingUnderstanding}), thereby shaping our understanding of model capabilities.
    Furthermore, we cannot yet rely on generative AI to produce Arabic visualisations. 
    For instance, ChatGPT currently generates charts using Python libraries like Matplotlib, which renders Arabic text incorrectly with disconnected letters (see \cref{fig:teaser}) unless additional Arabic text-shaping libraries (\eg, arabic-reshaper) are explicitly used. 
    We therefore invite the community to reconsider the weight afforded to defaults in the design process. Framing a configuration as the “default” implicitly positions alternatives as deviations from the norm, reinforcing dominant perspectives; supporting multiple contextualised defaults may offer a more equitable starting point for visualisation design. Including Arab practitioners in the design process for these tools is also essential.   
    Another implication of our work is to adopt D\"ork \etal's \cite{doerk2013CriticalVisualization} principles of critical visualisation, where creators could use the tooltip to show alternative place names, provide a disclosure describing the map's limitations, and hold the map's constructor accountable.

    %We also observed, through our interviews and tool analysis, that the power structure affects not only alignment and the position of the Y-axis in the default visualisation settings, but also country recognition and borders when designing maps. 
    %%link to previous literature
    %The discussion about how maps can be weaponised to show sympathy and importance is not new (e.g., the book "How to Lie with Maps"). We shed light on how single framing is embedded in most of these tools without notification or recognition of uncertainty.  

%\subsection{More user Research with Data Visualisation in Arabic}
\subsection{Research Agenda and Future Directions}
%% The claim: summarizes the paragraph
%% Evidences from our study
%% Connecting to previous literature
%% So what? the implication(s) or research agenda
While altering deeply rooted linguistic beliefs (\eg, ``designs in English are more elegant'') is difficult, interrupting this chain by developing empirically grounded design guidance for RTL scripts is more attainable.
Doing so requires empirical investigations that replicate core visualisation perception, interpretation and comprehension studies (\eg, \cite{Cleveland1984GraphicalMethods, quadri2022survey, tory2004human}) in RTL settings, and examine whether violating LTR conventions necessarily impairs comprehension, or whether such assumptions are themselves a product of normalised defaults.

This opens the door to many research opportunities. Mapping the visualisation literacy within audiences who consume Arabic script will provide an evidence-based understanding of their capabilities and strategies. 
Although robust measurements for data visualisation literacy exist (\eg, \cite{Lee2017VLAT:Test, Ge2023CALVI:Visualizations, firat2022interactive}), the empirical studies that ground them were often not designed with RTL scripts and their users in mind \cite{solen2022scoping}. 
Additionally, while data visualisation literacy at scale likely depends on other factors such as educational curricula, prior adaptation of visualisation literacy assessments \cite{andrianarivony2022investigating, omelchenko2025cross} provide concrete pathways for culturally and linguistically grounded measurement. %Designers may find these tools useful for getting a quick sense of their audience without making assumptions.
% Contextualized systematic assessments could therefore help designers understand how chart comprehension varies across different types of users 
Contextualising, adapting and validating existing visualisation literacy instruments would enable more precise claims about visualisation
skill distributions and help derive methods for teaching or improving data visualisation globally. 
    
Our understanding of the Arabic visualisation ecosystem, particularly through its users, remains limited. 
Our interview analysis shows that practitioners routinely make design choices based on assumptions about their audience's expectations for directionality rather than established guidelines or empirical evidence. Some even presume readers are accustomed to LTR charts.  
The lack of empirically tested guidelines for RTL data visualisation was surfaced in Alebri \etal's work
%  analysis of 128 visualisations from Arabic data journals, 
which revealed inconsistencies in how authors approach RTL designs and adaptations
% such as mirroring or axis reversal which appear ad hoc and rarely grounded in tested design principles
\cite{alebri2024design}.
Because \emph{guidelines} and \emph{tools} are largely shaped and evaluated through empirical observations \cite{engelke2018visupply, batziakoudi2025lost}, the lack of investigation for RTL contexts leaves practitioners without contextualised references. 
% For RTL contexts, directionality is one aspect of design that might oppose LTR. 
% While researchers have recommended obeying reading gravity, all aspects of the order in which the viewer perceives (reads) the visualisation \cite{sawicki2022visqualdex}, their claim mostly stems from evaluations of LTR experiences. 
Future research with Arabic visualisation audiences can address this gap by replicating core perception and interpretation studies to assess how mirroring affects visualisation comprehensions across chart types and user backgrounds.
Beyond directionality, future studies could also investigate how visualisation elements such as legends, annotations, title, and framing, interact with cultural conventions and shape people's attitude towards data visualisation \cite{peck2019dataa, kong2018frames, rakotondravony2023probablement}.
Overall, moving practitioner beliefs toward direct empirical evidence would allow researchers to determine when RTL adaptations improve comprehension, when they are neutral, and when they introduce new challenges. 
% And as guidelines iteratively emerge, codifying them into concrete, implementable design requirements can provide a foundation for more inclusive authoring tools, for example, with explicit RTL-aware defaults, or options that preserve semantic intent and formatting.

We provide actionable recommendations for visualisation authoring tools on how to benefit from our findings, particularly in restoring the creator's agency: 
(1) \textit{consistent design controls}: text formatting and layout options should remain available through discoverable controls with consistent terminology across tools, regardless of script direction, so that authoring in Arabic does not remove capabilities and interactivity that creators retain in LTR contexts;
(2) \textit{recognition of foundational elements}: Eastern numerals and Arabic city names should be handled as native data types rather than worked around;
% (\foreignlanguage{arabic}{٣}) \textit{decoupled and consistent support for directionality}: designers should be able to set the directions of text, axes, layouts through discoverable controls with consistent terminology across tools;
(3) \textit{transparency over silent overrides}: automatic transformation should be surfaced with option to revert. 
Together, these requirements would let RTL practitioners work within a unified, user-controllable set of design elements rather than continually negotiating around defaults.

As with all studies, some limitations should be noted. 
First, generalisation to other RTL scripts. 
Generalisation is problematic in itself, as it contributes to the constraints mentioned in  \cref{sec:working-under-constraints}. 
Therefore, our work may not apply to all RTL scripts; it uses Arabic as a case study, with a base of over two billion users \cite{w32023languages}, so there is a pressing need to consider them in the visualisation community \cite{Jena2021TheVisualization}. Focusing on a single script enables the development of tools for that script rather than adapting an existing tool. Therefore, we encourage explorations of other RTL scripts, such as Persian and Urdu. 
Second, the sample of participants. Some might argue that our pool is small, but given that data visualisation practice continues to flourish in the Arab world and that our participants are experienced and influential, we believe that quantity should be balanced by quality. We also recognise the underrepresentation of women in our sample. As described in \cref{sec:interview-method}, we have made a substantial effort to recruit a gender-balanced sample; this may be due to economic and cultural factors. 
Third, while there is a major intersection between the GUI-based tools analysed and the tools used by our interviewed participants, we recognise that some tools and other chart types need further assessment. Future work could inspect code-based tools more closely or evaluate the tools with more complex charts. %Our results, therefore, may not apply to other tools, frameworks, libraries, or charts. 
Lastly, our findings reflect the perspectives of native Arabic-speaking practitioners. However, non-Arabic-speaking practitioners also create visualisations for Arabic-script contexts, and their perspectives should therefore be considered in future research.      

%- Imbalanced sample of participants \\
%- If reviewer comments on generalization to other RTL $\rightarrow$ generalization IS the problem. 

%################################ SECTION ####################
\section{Conclusion}
Through an analytical evaluation of seven visualisation authoring tools with Arabic datasets and interviews with Arab practitioners, we explore the visualisation design practice for Arabic scripts. Our results suggest limited support for Arabic scripts in these tools, particularly in the recognition of Eastern numerals, requiring massive efforts from practitioners over foundational elements. Other observed issues related to text alignment, font options, Arabic country names on maps, and the lack of automatic mirroring features push creators to work under constraints, requiring them to adapt by using additional tools or settling for a basic chart or compromising interactivity. We argue that these limitations stem from a lack of agency, thereby disadvantaging creators of Arabic visualisations of the benefits of these tools. This also compromises their creativity. We contribute a reflection on infrastructure disproportion, by showing how defaults in visualisation tools are biased to LTR conventions, along with calls for action.

%designer literacy
%assumptions about the audience 

%% file: default-chart-table.tex
\definecolor{squareblue}{HTML}{666666}
\definecolor{squaregray}{HTML}{EEEEEE}
\definecolor{squarelightblue}{HTML}{6792F1}

\newcommand{\lgsq}{\raisebox{-0.3ex}{\textcolor{LightGreen}{\rule{\squaresize}{\squaresize}}}\xspace}

\setlength{\tabcolsep}{\hpad} % remove horizontal padding in tables
\renewcommand{\arraystretch}{1} % keep vertical spacing normal in tables

\newcommand{\gsqdiag}{%
  \raisebox{-0.3ex}{%
    \begin{tikzpicture}[baseline=(sq.base), inner sep=0pt, outer sep=0pt]
      \useasboundingbox (0,0) rectangle (\squaresize,\squaresize);
      \node[inner sep=0pt, anchor=south west] (sq) at (0,0)
        {\textcolor{squaregray}{\rule{\squaresize}{\squaresize}}};
      \draw[line width=0.4pt]
        ($(sq.center)+(-0.45*\squaresize,-0.45*\squaresize)$) --
        ($(sq.center)+(0.45*\squaresize,0.45*\squaresize)$);
    \end{tikzpicture}%
  }\xspace}

\begin{table*}[]
\centering
\caption{Arabic and RTL support in generating basic charts: \iconbar barchart, \iconline linechart, \iconmap map, using default settings in seven visualisation authoring tools. 
% \lgsq indicate tool support while \gsq refer to tool failure. 
RAWGraphs do not support maps. Criteria with binary values are presented in this table; the rest are in the supplementary material, along with the analysis using Western numerals. The tools collectively demonstrate insufficient support for Eastern numerals. }%, while Grey squares \gsq\ denote tool failure to support. }
\label{tab:default-chart-tool-analysis}
\begin{tabular}{rccccccccccccccccccccccccccc}
\multicolumn{1}{c}{} & \multicolumn{3}{c}{MS Excel} & & \multicolumn{3}{c}{Tableau} & & \multicolumn{3}{c}{Power BI} & & \multicolumn{3}{c}{Flourish} & & \multicolumn{3}{c}{Datawrapper} & & \multicolumn{3}{c}{Google Sheets}& & \multicolumn{3}{c}{RAWGraphs} \\
Chart type & \iconbar & \iconline & \iconmap & & \iconbar & \iconline & \iconmap & & \iconbar & \iconline & \iconmap & & \iconbar & \iconline & \iconmap & & \iconbar & \iconline & \iconmap & & \iconbar & \iconline & \iconmap & & \iconbar & \iconline \\
Arabic text rendering & \lgsq & \gsqdiag & \lgsq & 
& \lgsq & \gsqdiag & \lgsq & 
& \lgsq & \gsqdiag & \lgsq & 
& \lgsq & \gsqdiag & \gsq & 
& \gsq & \gsqdiag & \gsq & 
& \lgsq & \gsqdiag & \lgsq & 
& \gsq & \gsqdiag \\

Bi-directional text handling & \lgsq & \lgsq & \gsq & & \gsq & \gsq & \gsq & & \lgsq & \lgsq & \lgsq & & \lgsq & \lgsq & \lgsq & & \gsq & \gsq & \lgsq & & \gsq & \lgsq & \gsq & & \gsq & \gsq \\
Eastern number rendering & \lgsq & \lgsq & \lgsq & & \gsq & \gsq & \gsq & & \gsq & \gsq & \gsq & & \gsq & \gsq & \gsq & & \gsq & \gsq & \gsq & & \gsq & \gsq & \gsq & & \gsq & \gsq \\ 

\multicolumn{28}{l}{} \\
\cline{1-1} \\[-8pt]
\multicolumn{28}{l}{\lgsq and \gsq indicate the presence and absence of tool support, respectively. } \\
\multicolumn{28}{l}{\gsqdiag indicates that the default line chart did not include textual data for evaluation.}\\

\end{tabular}
\end{table*}